\documentclass[12pt]{article}
\usepackage{graphicx}
\begin{document}
\bigskip
\centerline{\Large \bf Simulation of Consensus Model of Deffuant et al}

\medskip
\centerline{\Large \bf on a Barab\'asi-Albert Network}

\bigskip
D. Stauffer$^1$ and H. Meyer-Ortmanns$^2$

\bigskip
Laboratoire PMMH, Ecole Sup\'erieure de Physique et Chimie
Industrielle, 10 rue Vauquelin, F-75231 Paris, France

\medskip

\noindent
$^1$ Visiting from Institute for Theoretical Physics, Cologne
University, D-50923 K\"oln, Euroland; stauffer@thp.uni-koeln.de

\noindent
$^2$ Visiting from School of Engineering and Science,
International University Bremen, P.O.Box 750561,
D-28725 Bremen, Germany; h.ortmanns@iu-bremen.de 

\bigskip

Abstract: In the consensus model with bounded confidence, studied
by Deffuant et al. 
(2000), two randomly selected people who differ not too much in their opinion
both shift their opinions towards each other. Now we restrict
this exchange of information to people connected by a scale-free network. As
a result, the number of different final opinions (when no complete consensus 
is formed) is proportional to the number of people.

\medskip

\noindent Keywords: Scale free networks, sociophysics, opinion formation,
Axelrod model.

\section{Introduction}

Partially motivated by a model of Axelrod \cite{Axelrod}, 
the model of Deffuant 
et al. \cite{Deffuant} simulates the building of a consensus, or the lack of
consensus, out of many initially random opinions. Other consensus models are 
those of Krause and Hegselmann \cite{Hegselmann}, Sznajd \cite{Sznajd} (for 
a review see \cite{Stauffer}), and Galam \cite{Galam}; they were summarized
recently in \cite{sociocise} and are part of sociophysics \cite{Galam} or 
sociodynamics \cite{Weidlich}, which belong to the wider field of 
interdisciplinary applications of statistical physics methods \cite{book}.

The Deffuant model is the one where the largest number of people was
simulated so far \cite{sociocise} so that the statistics was best
and this is the reason why we selected this model for the present study.
It assumes that everybody can talk with everybody else
with the same probability,
similarly to random graphs, but with sites living in the continuum. (So the 
model was not considered on random graphs in the sense of Erd\"os and 
R\'enyi \cite{reni}.)
In this unrealistic limit, analytical approximations
work well \cite{Ben-Naim}. The opposite limit of people restricted to a square
lattice was also simulated, with interaction between close neighbours (like
nearest neighbours) only, but 
may apply better to trees in an orchard than to human beings. Real social
connections may lie in between, with few people having lots of friends 
and many people having few friends to talk with. Everybody is still connected
with everybody but only indirectly over a short link of mutual friends. 
The best
studied model for these types of connections are the scale-free networks 
of Barab\'asi and Albert \cite{BA} where the number of people having $k$
friends decays as $1/k^3$.

The effect of network topologies on the dissemination of culture \cite{Axelrod}
or on the spreading of information will be studied in future work \cite{hmo}.

The next section defines the two models, with directed and undirected 
bonds in a Barab\'asi-Albert network, while section 3 gives the results
and section 4 the conclusions.

\section{Models}
The Barab\'asi-Albert network starts with a small number $m$ ($m=3$ in our 
simulations) of sites (agents, people) all connected with each other. 
(We varied $m=3$ also to $m=2$, $4$ and $5$ and observed similar test
results for 100 runs each.) Then a 
large number $N$ of additional sites is added as follows: Each new site 
selects $m$ of the already existing sites as friends, with a probability 
proportional to the number of friends this already existing site had before.
When the new site A has selected an already existing site B as friend, this
selection increases for both A and B the number of friends by one. In the 
usual undirected Barab\'asi-Albert model, later A can talk with B and B 
can talk with A. In the simpler directed version, A initiates a talk with B,
but B initiates talks only with those $m$ people whom B had selected as
friends. Thus in this directed version, everybody has a fluctuating number of
people connected with him,
but asks only one of exactly $m$ people for advice at a time,
and these are the people the new site had selected when joining the network.
(One may think of a hierarchy of bosses and underlings.) For the undirected
case no such distinction between friends and people to talk with is needed; the
connection network of friends then is constructed as in \cite{aa}.

Once the network has been constructed, we start the consensus process of
Deffuant et al.. Everybody gets a random number $S$ between zero and one as
initial opinion. Then for each iteration, every site A is updated once by
selecting randomly one site B from the sites connected with A. In the undirected
case the selection is taken from all $k$ sites who had selected A as friend
or whom A has selected as friends. In the directed case the selection is made 
only from the $m$ sites which A had selected as friends. If then the opinions 
$S_A$ and $S_B$ differ by more than a constant confidence bound $\epsilon$ 
between zero and one, A and B refuse to discuss and do not change their 
opinion. 
Therefore $\epsilon$ may be interpreted as a measure for the
tolerance of people to other opinions.
Otherwise both move closer to the position of the other by an 
amount $\delta = \mu (S_A - S_B)$ with $\mu = 0.3$ in our simulations, i.e. A 
takes the opinion $S_A - \delta $ and B the opinion $S_B + \delta $. The
parameter $\mu$ characterizes the flexibility in changing the opinion. After
sufficiently many iterations (unfortunately much more than the $\sim 10^2$
in the ``random'' network \cite{Deffuant}) no opinion moves by more than 
$2 \times 10^{-8}$: a fixed point in the space of opinions
is approximated. (For different $\mu$ this small value needs to be
adapted.) Hundred samples of this type were averaged over.
The opinions are then
placed in bins of width $10^{-6}$ and are counted by checking which bins
are occupied and do not have the lower neighbouring bin occupied. In this
way, the total number of fixed opinions is found.
The ``directed'' Fortran program is available from the authors.

\begin{figure}[hbt]
\begin{center}
\includegraphics[angle=-90,scale=0.5]{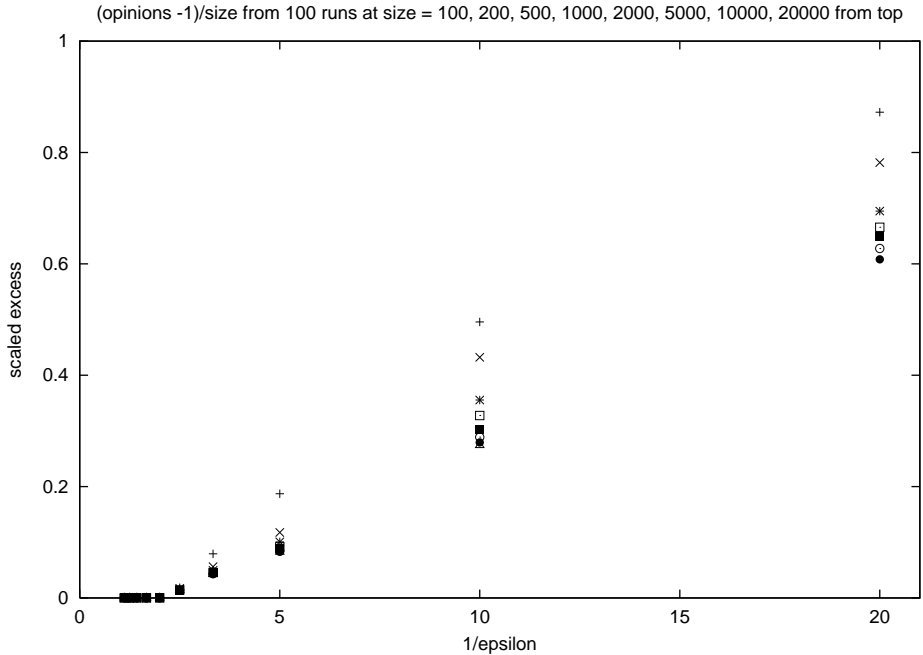}
\end{center}
\caption{
Scaled excess number $F_E = (F-1)/N$ of different opinions in the undirected
network, with $N$ between 100 and 20000 given in the headline.
}
\end{figure}

\begin{figure}[hbt]
\begin{center}
\includegraphics[angle=-90,scale=0.5]{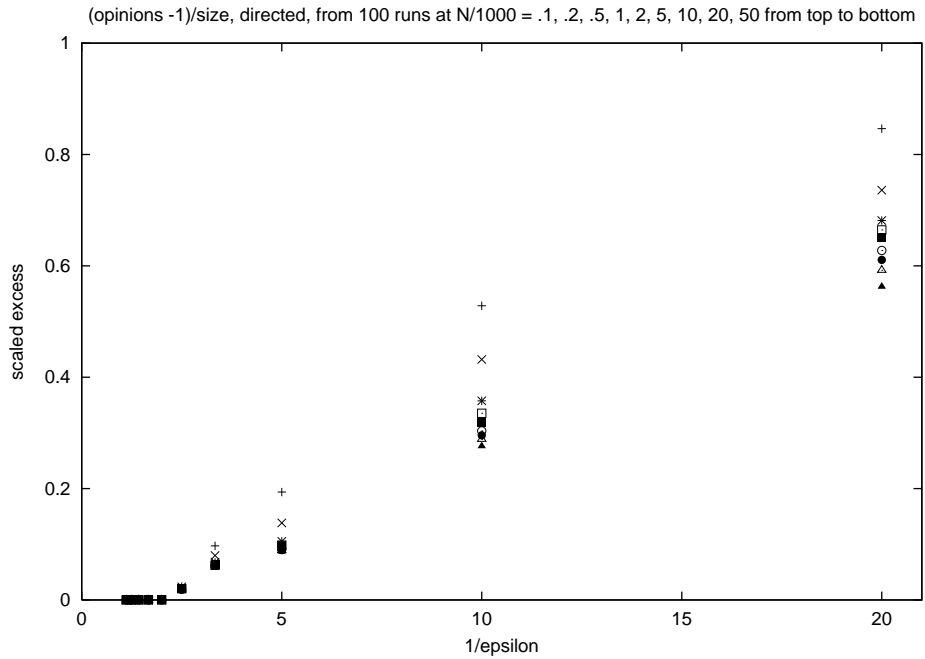}
\end{center}
\caption{
As in Fig.1, but for the directed case.  
}
\end{figure}

\begin{figure}[hbt]
\begin{center}
\includegraphics[angle=-90,scale=0.5]{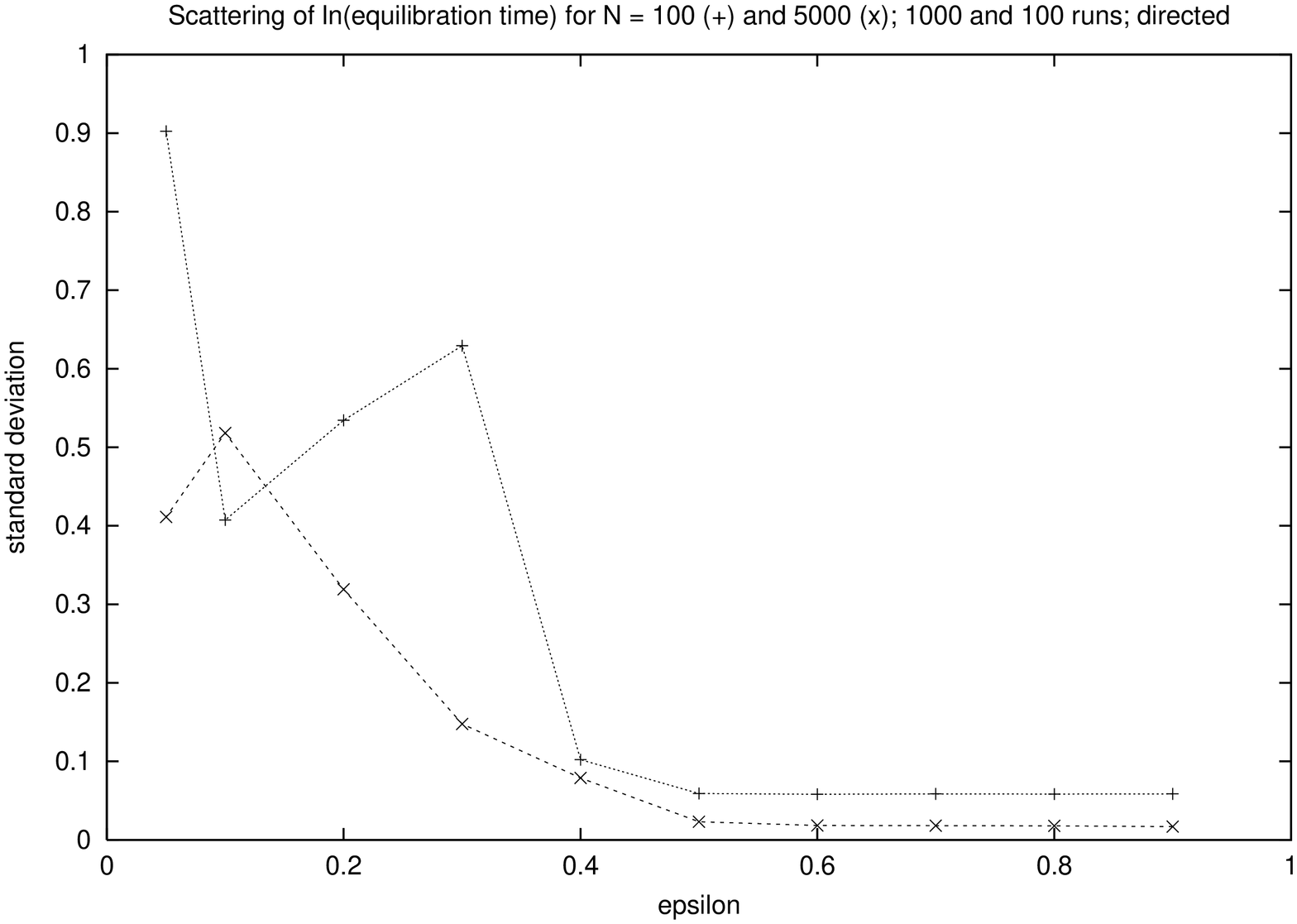}
\end{center}
\caption{
Fluctuations in the time $t_c$ to reach fixed points, in Monte Carlo steps per 
site, for $N = 100$ and 5000 for the directed case. We plot the standard 
deviation $\sigma$ of the logarithm, defined through $\sigma^2 = <(\ln t_c)^2>
- <\ln t_c>^2$.
}
\end{figure}

\section{Results}

Figure 1 shows the undirected and figure 2 the directed case. For $\epsilon$
larger than about 0.4 a full consensus is reached; only one opinion survives.
For smaller $\epsilon$, no consensus is reached and the number $F$ of fixed 
opinions increases with decreasing $\epsilon$. When the number $N$ of people
increases, the $F$ for small $\epsilon$ also increases $\propto N$ for large
$N$. This increase is the crucial difference to the random version without
Barab\'asi-Albert restriction, when $F$ is independent of $N$ for large $N$.
Thus we plot in figures 1 and 2 the scaled excess number
$$ F_E = (F-1)/N$$
versus $1/\epsilon$: in the random case in the sense of Deffuant et al.
(without specified network topology) $F$ roughly equalled
$1/\epsilon$, while now $F \sim N/\epsilon$. 
The remaining mild N-dependence of the scaled excess number which is seen
in Figs. 1 and 2 is interpreted as a finite size effect getting weaker for 
larger system sizes $N$, cf. the figures.

Figure 3 shows the enormous fluctuations in the number $t_c$ of iterations 
needed to reach the fixed point, similarly to the Snajd model \cite{Sznajd}. 
This feature is understood from the fact that 
reaching a certain fixed point, e.g. that of consensus, is a collective 
property of the $N$ agents which cannot be obtained as an average over 
subsystems. Values of such kind of collective quantities may strongly 
fluctuate like those of individual ones, even for $N \rightarrow \infty$,
i.e. they are not self-averaging. However, the behaviour near $\epsilon = 0.1$
for $N = 5000$ is not understood.

\section{Conclusions}

Our condition for the confidence bound $\epsilon > 0.4$ to allow a complete 
consensus is about the same on the Barab\'asi-Albert network as it was
in the usual random case \cite{Deffuant}. But when no consensus is formed
because $\epsilon < 0.4$ is too small, then our number of different opinions is
proportional to the number of people involved and no longer size-independent.  
This observation shows that the choice of the links between the agents may be 
crucial in studies of social behaviour.
\bigskip

\noindent {\bf Acknowledgements}: We thank PMMH at ESPCI for the warm 
hospitality, Sorin T\u{a}nase-Nicola for helping us with the computer 
facilities, and G. Weisbuch for drawing out attention to Axelrod's model.

\end{document}